\documentclass[final,numcites,sort&compress]{aipproc}

\layoutstyle{6x9}

\begin{document}

\title{Proton-rich nucleosynthesis and nuclear physics}


\author{T. Rauscher}{
  address={Dept.\ of Physics, University of Basel, 4056 Basel, Switzerland}
}

\author{C. Fr\"ohlich}{
  address={Dept.\ of Physics, NCSU, Raleigh, NC 27695, USA}
}


\begin{abstract}
Although the detailed conditions for explosive nucleosynthesis are derived from astrophysical modeling, nuclear physics determines fundamental patterns in abundance yields, not only for equilibrium processes. Focussing on the $\nu$p- and the $\gamma$-process, general nucleosynthesis features within the range of astrophysical models, but (mostly) independent of details in the modelling, are presented. Remaining uncertainties due to uncertain $Q$-values and reaction rates are discussed.
\end{abstract}

\maketitle


\section{Introduction and general considerations}

Several nucleosynthesis processes have been suggested to include proton-rich nuclei under explosive conditions \cite{tomnic,pprocreview}: rapid proton captures (the rp-, $\nu$p-, np-process) and photodisintegrations (the $\gamma$-process). Although the detailed conditions for explosive nucleosynthesis are derived from astrophysical modeling, nuclear physics determines fundamental patterns in abundance yields, not only for equilibrium processes. While the hydrodynamics sets the range of feasible conditions and the timescale of the process, nuclear structure (e.g., through reaction $Q$-values) as well as reactions constrain the nucleosynthesis path and limit the achievable abundances. More specifically, astrophysical conditions may determine how far a nucleosynthesis path extends but nuclear physics determines the location of, and the relative abundances of the nuclei produced along, the path. Moreover, nuclear structure poses limits on how far nucleosynthesis can proceed even with extended timescales (e.g., in the rp- and $\nu$p-processes). Nuclear properties also determine which reactions will be important. Again, this is connected to the $Q$-values but also to Coulomb barriers which have to be overcome.

In many cases, nuclear physics already constrains a given nucleosynthesis process to a comparatively small range of conditions. Outside this range, either no appreciable change in abundance occurs or the nuclei are destroyed completely. Therefore, it is possible to construct an effective process path and this allows trajectory-independent considerations of the impact of nuclear uncertainties within the permitted range of conditions. It turns out that the locations of these uncertainties (nuclei, involved rate types) are quite robust.

Here, we briefly discuss two examples and their related nuclear uncertainties. For further details, see \cite{frorautru} for the $\nu$p-process and \cite{rau06,rapp06,pprocreview} for the $\gamma$-process. In both cases, we make use of
\emph{rate field plots}, as shown in Figs.\ \ref{fig:equiabuns}--\ref{fig:gammalo}, which are extremely helpful to examine a nucleosynthesis process. The shade of each nucleus in the plot is either its relative abundance within an isotopic chain (Fig.\ \ref{fig:equiabuns}) or related to its lifetime with respect to the fastest reaction destroying it. The arrows give the direction of the fastest \emph{net rate per target nucleus}. The net rate is obtained from the difference of a rate per target nucleus and its reverse rate. These are not the actual rates in a reaction network as these would also depend on the abundance of the target nucleus but they allow to quickly gather which reaction dominates on a nucleus and which direction the synthesis path would take, were the nucleus actually produced.

\section{The $\nu$p-process}
\begin{figure}
  \includegraphics[width=0.65\textwidth]{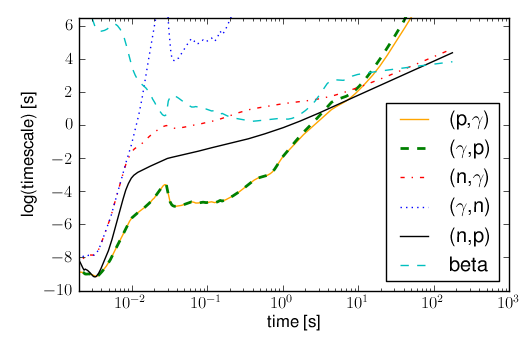}
  \caption{\label{fig:timescales}Average lifetime comparison showing (p,$\gamma$)-($\gamma$,p) equilibrium.}
\end{figure}

The innermost, still ejected layers of core-collapse supernovae become proton-rich due to the interaction with the intense $\nu$-wind within which they move outwards. The material quickly cools from the initially high temperature, assembling nucleons mainly to $^{56}$Ni and $\alpha$-particles, leaving a large number of free protons. At sufficiently low temperature, rapid proton captures ensue on $^{56}$Ni. Production of heavier nuclei would not be possible without the tiny fraction of free neutrons which is created by $\overline{\nu}_\mathrm{e}$ captures on free protons. The supply of free neutrons allows (n,p) reactions bypassing slow electron captures and $\beta^+$ decays (see Fig.\ \ref{fig:timescales}) and thus providing the possibility of sequences of proton captures and (n,p) reactions producing nuclei with larger and larger $Z$ and $A$. This was termed the $\nu$p-process (for further details, see, e.g.,  \cite{frothis} in this volume, and \cite{fro06a,fro06b,pruet06,wanajo11}).

\begin{figure}
  \includegraphics[width=1.2\textwidth]{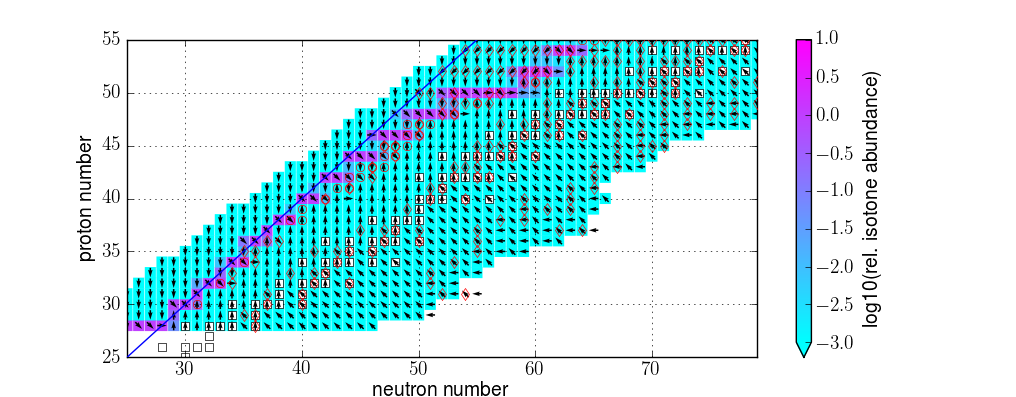}
  \caption{\label{fig:equiabuns}Equilibrium abundance in each isotonic chain (color-shaded), dominating reactions (arrows), and nuclear uncertainties (circles and diamonds). Stable nuclei are marked as white squares.}
\end{figure}
\begin{figure}
  \includegraphics[width=1.2\textwidth]{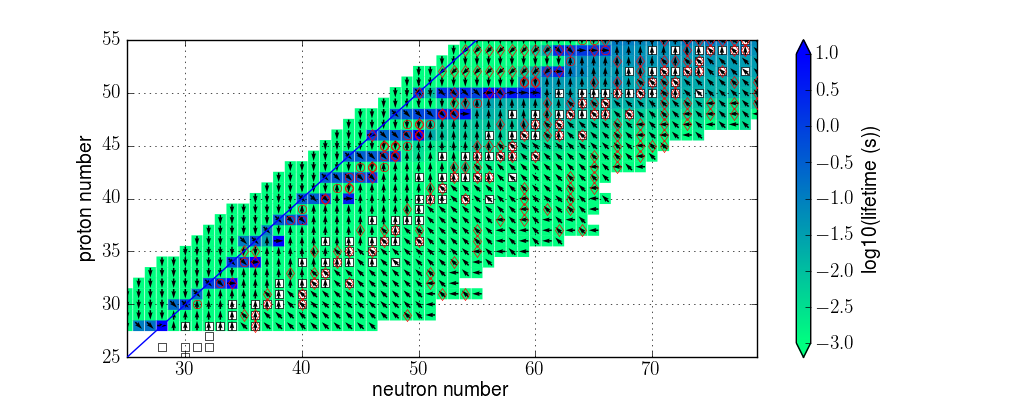}
  \caption{\label{fig:flux}Same as Fig.\ \ref{fig:equiabuns} but the shade of each nucleus gives its lifetime against the dominating destruction reaction.}
\end{figure}

\begin{figure}
  \includegraphics[width=1.2\textwidth]{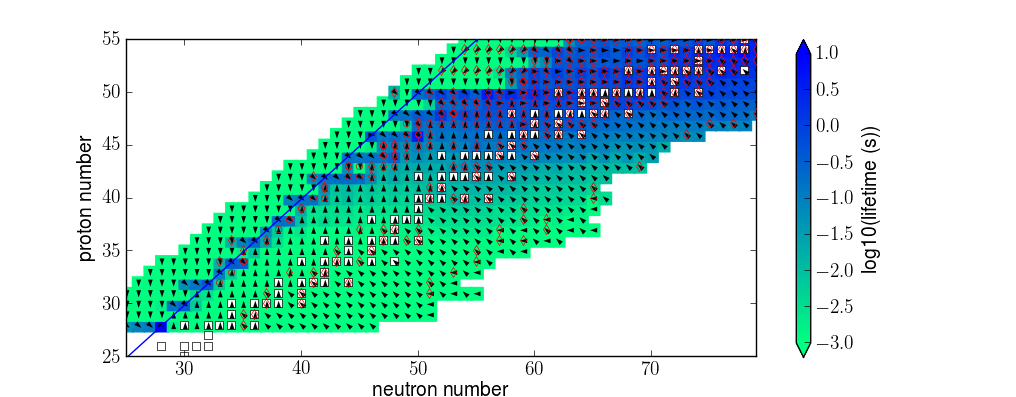}
  \caption{\label{fig:lateflux}Same as Fig.\ \ref{fig:flux} but at very late times; neutron captures drive the path even further to stability.}
\end{figure}

The location of the effective $\nu$p-process path remains remarkably unaffected by variations of the astrophysical parameters, such as entropy, expansion timescale, and details of the reverse shock \cite{frorautru,frothis,fro06b,pruet06,wanajo11}. Also systematic variations of reaction rates show only small effects, if any, regarding the path location \cite{frorautru,frothis,wanajo11}. All these variations, however, determine how far up the path is followed or whether it is terminated already at low $Z$. Thus, also the achieved abundances within the path are determined by these conditions. The reason for this behavior can easily be understood when realizing that (p,$\gamma$)-($\gamma$,p) equilibrium is upheld until late times as shown in Fig.\ \ref{fig:timescales}. In consequence, the abundance maximum in each isotonic chain is given by the equilibrium abundance, color-shaded for each nucleus in Fig.\ \ref{fig:equiabuns}. The corresponding Fig.\ \ref{fig:flux} shows the same rate field but with the lifetimes of the nuclei. Due to the (p,$\gamma$)-($\gamma$,p) equilibrium each proton-rich nucleus in an isotonic chain is rapidly (compared to the expansion timescale) converted into the nucleus favored by the equilibrium conditions. Therefore, the maximum within a chain is characterized by a low (p,$\gamma$) reaction $Q$-value because the relative speed of forward and reverse rate depends exponentially on the $Q$-value \cite{frorautru,tomreview}. The largest flux into the next isotonic chain occurs at these nuclei, which would be waiting points in a pure rp-process \cite{schatz}. As can be seen in Fig.\ \ref{fig:flux}, these nuclei indeed exhibit longer lifetimes but they may still be overcome if the neutron abundance is sufficiently high. This implies that a variation of the neutron density or the (n,p) rate on these waiting points will mostly affect how fast nuclei with larger $Z$ can be reached within the timescale given by the expansion.

Another nuclear-structure determined feature can be seen immediately in Figs.\ \ref{fig:equiabuns} and \ref{fig:flux}: while the waiting points follow the $Z=N$ line up to Mo, they are extending further and further to neutron-richer isotopes between Mo and Sn, pushing the path gradually away from the $Z=N$ line. The path is pushed strongly towards stability at the Sn isotopes and above, providing a strong barrier for the efficient production of any elements beyond Sn. Two effects are acting here: The location of the waiting-points at neutron-richer nuclides with longer lifetimes with respect to both decays and (n,p) reactions, and the fact that also the proton captures become slower due to the higher Coulomb barriers at larger $Z$. The latter leads to a dominance of (n,$\gamma$) reactions over (p,$\gamma$) and ($\gamma$,p) as can be seen in Figs.\ \ref{fig:equiabuns} and \ref{fig:flux} at high $Z$, and especially in Fig.\ \ref{fig:lateflux} showing the situation at late times. The neutron captures push the nucleosynthesis strongly towards stability and even beyond and prevent the production of appreciable amounts of matter above the Sn region.

In Figs.\ \ref{fig:equiabuns}--\ref{fig:lateflux} circles mark nuclei where remaining experimental uncertainties in the $Q$-values (mass differences) affect the $\nu$p-process results. The diamonds mark targets for which (p,$\gamma$), (n,p), or (n,$\gamma$) rates are close and theoretical uncertainties in the predicted rates may be important. These occur only at large $Z$, reached at late time if at all. Additionally, (n,p) reactions on all waiting points shown in Fig.\ \ref{fig:equiabuns}, and determining the flow to heavier nuclei, require further investigation. For a detailed list, see \cite{frorautru}.

The above general conclusions hold for any trajectory yielding an appreciable $\nu$p-process. Moreover, the range of conditions permitting such a process are also set by nuclear properties. Without neutrons, the waiting points cannot not be passed but a large neutron density (caused, e.g., by a higher $\overline{\nu}_\mathrm{e}$ flux) would make the (n,p) reactions faster than (p,$\gamma$) on \emph{any} proton-rich nuclide above $^{56}$Ni. On the other hand, too high a temperature prevents the outbreak from the $^{56}$Ni seed as ($\gamma$,p) and ($\gamma$,$\alpha$) reactions are faster than proton captures and always establish QSE with $^{56}$Ni. Finally, in a freeze-out charged-particle captures are suppressed according to the Coulomb barriers and a further increase in $Z$ is more and more hampered. The exact values determining the window of favorable conditions can thus be estimated from nuclear properties in the spirit of the classical B$^2$FH paper \cite{b2fh} (see \cite{frorautru} for details).
The astrophysical hydrodynamics just determines how long the ejected matter is subjected to those favorable conditions and thus how far up in $Z$ and $A$ $\nu$p nucleosynthesis can proceed.

\section{The $\gamma$-process}

\begin{figure}
  \includegraphics[width=1.2\textwidth]{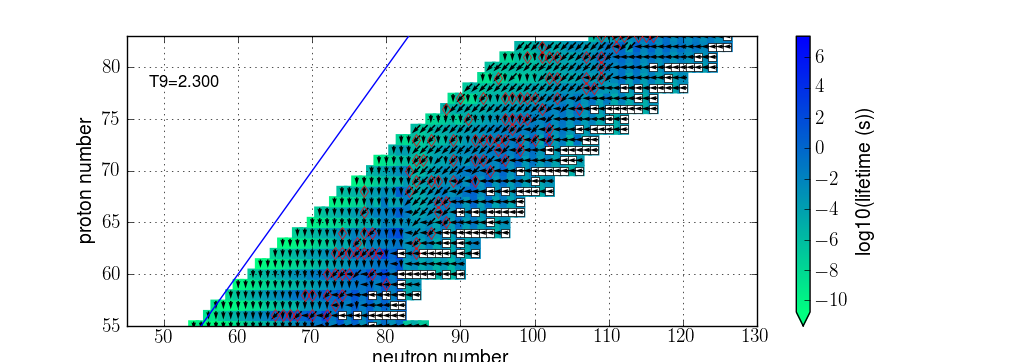}
  \caption{\label{fig:gammahi}Same as Fig.\ \ref{fig:flux} but for the $\gamma$-process at $T=2.3$ GK.}
\end{figure}
\begin{figure}
  \includegraphics[width=1.2\textwidth]{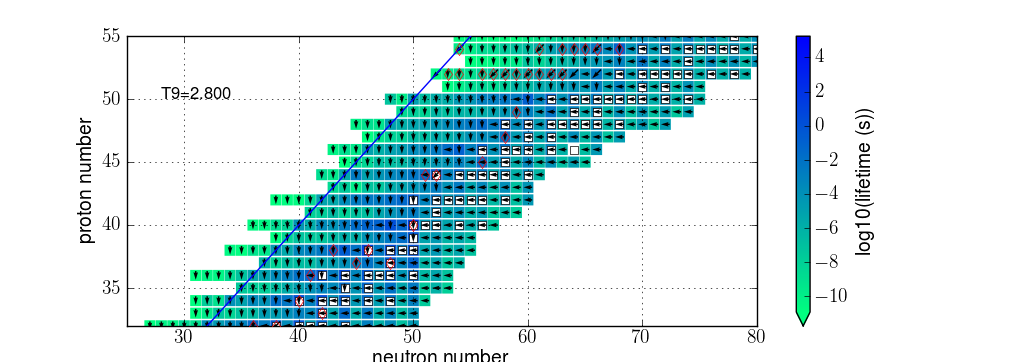}
  \caption{\label{fig:gammalo}Same as Fig.\ \ref{fig:gammahi} but for lighter elements at $T=2.8$ GK.}
\end{figure}

Photodisintegration of matter in the \emph{outer} shells, more specifically the O/Ne layers, of a massive star exploding by a core-collapse supernova has been termed $\gamma$-process and proposed to be the source of the p-nuclei \cite{woohow,ray95}. Alternative sites include type Ia supernovae \cite{meywoo,travag} but, although the nuclear seed distribution is different, the photodisintegration mechanism remains similar. Although the $\gamma$-process does not involve reaction equilibria, again nuclear structure determines the reaction flow and similar reasoning as for the $\nu$p-process can be applied.

Figures \ref{fig:gammahi} and \ref{fig:gammalo} show the rate fields in two sections of the nuclear chart, for two characteristic temperatures. Starting at stability (white squares) the typical $\gamma$-process pattern is found, an initial sequence of ($\gamma$,n) reactions until the path is deflected to lower $Z$ by ($\gamma$,p) or ($\gamma$,$\alpha$). At first glance it is obvious that ($\gamma$,$\alpha$) deflections dominate in the higher mass range (Fig.\ \ref{fig:gammahi}) whereas ($\gamma$,p) dominate below $N=82$ (Figs.\ \ref{fig:gammahi}, \ref{fig:gammalo}). This is due to the distribution of proton and $\alpha$ separation energies, determining the $Q$-values of ($\gamma$,p) and ($\gamma$,$\alpha$), and thus depends on the nuclear masses. This also explains the findings of \cite{rau06,rapp06} regarding the different sensitivity of the production of p-nuclei to ($\gamma$,p) and ($\gamma$,$\alpha$) variations at lower and higher mass.

As above, competition points of rates are marked by open diamonds in Figs.\ \ref{fig:gammahi} and \ref{fig:gammalo}. (There are no mass uncertainties because there are no equilibrium waiting points and all masses in the $\gamma$-process range are known with high accuracy). These are nuclei for which the rates for ($\gamma$,n) and/or ($\gamma$,p) and/or ($\gamma$,$\alpha$) are close to each other and it is undecided within the theoretical uncertainty which of the reactions dominates. Although a large number of such competition points is displayed, it has to be remembered that only the first one or two, counted from stability, within an isotopic chain are relevant as the $\gamma$-process commences with photodisintegration of stable nuclides.


\begin{theacknowledgments}
This work was partially supported
by a DOE topical collaboration, contract DE\_FG02-10ER41677,
by the European Commission within the FP7 ENSAR/THEXO project, and by the ESF through the EuroGENESIS program.
\end{theacknowledgments}



\bibliographystyle{aipproc}   


\end{document}